\documentclass{elsarticle}
\usepackage{lineno,hyperref}
\usepackage[group-separator={,}]{siunitx}
\usepackage{multicol}
\usepackage{graphicx}
\usepackage{fancyhdr}
\usepackage{amsmath}
\usepackage{amsfonts}
\usepackage{multirow}
\usepackage{adjustbox}
\usepackage{epstopdf}
\usepackage{subcaption}
\usepackage{amsmath}
\usepackage{calrsfs}

\journal{}

\begin{document}
	
	\begin{frontmatter}
		
		\title{Breast simulation pipeline: from medical imaging to patient-specific simulations}
		
		\author[mymainaddress]{Arnaud Mazier}
		\ead{mazier.arnaud@gmail.com}
		
		\author[mymainaddress]{Stéphane P.A. Bordas\corref{mycorrespondingauthor}}
		\cortext[mycorrespondingauthor]{Corresponding author}
		\ead{stephane.bordas@me.com}
		
		\address[mymainaddress]{Institute of Computational Engineering, Department of Engineering, Université du Luxembourg, Esch-sur-Alzette, Luxembourg}
		
		\begin{abstract}
			\noindent\textit{Background} 
			Breast-conserving surgery is the most acceptable operation for breast cancer removal from an invasive and psychological point of view. Before the surgical procedure, a preoperative MRI is performed in the prone configuration, while the surgery is achieved in the supine position. This leads to a considerable movement of the breast, including the tumor, between the two poses, complicating the surgeon's task.
			\medbreak
			
			\noindent \textit{Methods} In this work, a simulation pipeline allowing the computation of patient-specific geometry and the prediction of personalized breast material properties was put forward. Through image segmentation, a finite element model including the subject-specific geometry is established. By first computing an undeformed state of the breast, the geometrico-material model is calibrated by surface acquisition in the intra-operative stance.
			\medbreak
			
			\noindent \textit{Findings} Using an elastic corotational formulation, the patient-specific mechanical properties of the breast and skin were identified to obtain the best estimates of the supine configuration. The final results are a Mean Absolute Error of \SI{4.00}{mm} for the mechanical parameters $E_{\mathrm{breast}} = 0.32$ \si{kPa} and $E_{\mathrm{skin}} = 22.72$ \si{kPa}, congruent with the current state-of-the-art. The Covariance Matrix Adaptation Evolution Strategy optimizer converges on average between $5$ to $30$ \si{min} depending on the initial parameters, reaching a simulation speed of \SI{20}{s}. To our knowledge, our model offers one of the best compromises between accuracy and speed.
			\medbreak
			
			\noindent \textit{Interpretation} Satisfactory results were obtained for the estimation of breast deformation from preoperative to intra-operative configuration. Furthermore, we have demonstrated the clinical feasibility of such applications using a simulation framework that aims at the smallest disturbance of the actual surgical pipeline.
			\medbreak
			\noindent\textbf{Word count:} 250 (abstract) and 3856 (paper).
		\end{abstract}
		
		
	\end{frontmatter}
	

	\section{Introduction}
	In breast surgical practice, images are acquired in the prone configuration, whereas surgery occurs in the supine position~\cite{Mazier2021}. We focus on using the Finite Element (FE) method to simulate the deformation of the breast from one stance to another to predict the final position of the tumor. The model behavior is driven by the rheological parameters of the breast, which can significantly modify the final breast shape as well as the (highly variable) breast anatomy. Despite remarkable progress~\cite{Sutula2020, Briot2022-2}, breast parameters are often challenging to measure \emph{in-vivo} and can evolve in time, depending on several biological factors. 
	
	Over the past decades, biomechanical breast models have been extensively studied for various medical applications such as surgical training, pre-operative planning, image registration, or tumor propagation~\cite{Chung2008, Han2011, Rajagopal2010, Samani2001, Ruiter2004, Sturgeon2016, Lavigne2022-2, Alcaniz2022}. Several studies relied on FE models to describe the geometry of the breast's organs (ligaments, skin, adipose, mammary gland, and fascia tissues) and the complexity of their behaviors. Henceforth, before selecting a model, a careful reflection has to be accomplished concerning the anatomical structure details and the required complexity (fidelity) to describe such behaviors. Contrary to FE mammogram simulations, breast-conserving surgery involves a larger spectrum of deformations requiring a precise anatomical model~\cite{Rebecca2022, Liu2017, Salmon2017, Mira2018-2}. For instance, when simulating the change in gravity direction from prone to supine, the breast totally changes shape from a hemisphere to an almost flat organ~\cite{Georgii2016, Rajagopal2007-2, Rajagopal2010}. A specific requirement is to reach a suitable accuracy compared with clinical data. Therefore, such reasonable precision can solely be achieved through personalized simulations aiming at tuning the mechanical and anatomical breast properties.
	
	In the context of breast preoperative surgical planning, extensive work has been conducted, emphasizing different features.
	
	\begin{itemize}
		\item Data acquisition. In the breast-conserving surgical pipeline, MRI is the imaging modality for pre-operative screening. However, most studies resorted to MRI~\cite{Rajagopal2007, Rajagopal2007-2, Han2011, Han2014, Eiben2016, Mira2018, BabarendaGamage2012, BabarendaGamage2019} or CT acquisitions in intra-operative configurations for inferring patient-specific mechanical properties. 
		
		\item Number of patients. As mentioned previously, gathering validation data is challenging as those imaging routines are not part of the surgical pipeline and require extra steps for the surgeon. At the time of writing, in the literature, the number of patients can vary from 1~\cite{Mira2018} to 86 patients~\cite{BabarendaGamage2012, BabarendaGamage2019}.
		
		\item Breast heterogeneity. The breast is a complex organ made of several sub-entities, such as the skin, mammary gland, adipose tissues, and ligaments leaning on the pectoral muscle. The most detailed model differenciated the left and right breast properties to attain the following properties $E^{right}_{\text{breast}} = \SI{0.3}{kPa}, E^{left}_{\text{breast}} = \SI{0.2}{kPa}, E_{\text{skin}} = \SI{4}{kPa}, E_{\text{fascia}} = \SI{120}{kPa}, E_{\text{ligaments}} = \SI{120}{kPa}, E_{\text{muscle}} = \SI{10}{kPa}$~\cite{Mira2018}, in firm agreement with the literature. Conversely, other models simplified the breast complexity using only a sole material to model the whole breast behavior, obtaining breast properties ranging on average from $0.25$ to \SI{0.3}{kPa}~\cite{BabarendaGamage2012, BabarendaGamage2019}. 
		
		\item Model. Several study modeled the breast using a Neo-Hookean material accounting for the non-linearities of the breast~\cite{Rajagopal2007, Rajagopal2007-2, Carter2009, Han2011, Han2014, Eiben2016, Mira2018, BabarendaGamage2012, BabarendaGamage2019}.
		
		\item Boundary conditions. The connection between the pectoral muscle and the breast is the unique boundary condition to model. Several studies only considered natural Dirichlet boundary conditions (null displacement, the inner breast is attached to the pectoral muscle)~\cite{Rajagopal2007, Rajagopal2007-2, Carter2009, Eiben2016}. While others opted for Neumann boundary conditions (sliding between the inner breast and the pectoral muscle)~\cite{Han2011, Han2014, Mira2018, BabarendaGamage2012, BabarendaGamage2019}. Therefore, overconstrained boundary conditions such as natural Dirichlet might significantly impact soft material behavior estimation. Conversely, authorizing too much freedom in the boundary conditions (sliding only) can lead to stiff material properties. 
		
		\item Error. Among the literature, different measurements and metrics have been used. For instance, few studies focused solely on anatomical landmarks (e.g., nipples) measurements~\cite{Carter2009, Han2011, Han2014, Eiben2016}, while a vast majority investigated specific metrics representing the overall error over the breast. Among the different global metrics,~\cite{Rajagopal2007, Rajagopal2007-2} predicted the supine stances of two patients with a surface RMS (Root Mean Square) error of \SI{8.4}{mm} and \SI{7.7}{mm}. Then,~\cite{Mira2018} calculated Hausdorff distances between the estimated and measured breast geometries for prone, supine, and supine tilted configurations equal to \SI{2.17}{mm}, \SI{1.72}{mm}, and \SI{5.90}{mm} respectively. Finally,~\cite{BabarendaGamage2012, BabarendaGamage2019} reported an RMS error of \SI{5}{mm} over 86 patients from prone to supine stance.
	\end{itemize}

	Despite several differences in the modeling approach of the above studies, some intersections are discernable. Firstly, they all employed a Neo-Hookean model with Young's modulus of the breast comprised between \SI{0.2}{kPa} and \SI{0.4}{kPa}. Secondly, even if different metrics were used, the simulation error from prone to supine stance ranged from $1.90$ $\pm$ \SI{2.17}{mm} to \SI{5}{mm} using mean and RMS errors, respectively. Therefore, most studies relied on MRI or CT data in the intra-operative configuration, which is incompatible with the current breast-conserving surgery pipeline. Despite adequate accuracy, few studies actually attempted to simulate the tumor movements and solely focused on breast deformation, making a relevant clinical application problematic. Finally, a few investigations communicated the run-time of the overall procedure that has to be compatible with clinical time scale; ideally, less than \SI{20}{min} between the imaging and the surgery.
	
	In this paper, we will present a new, fast, and simple patient-specific FE pipeline addressing the challenge of estimating the intra-operative configuration from clinical imaging acquisition. We will first introduce the different steps to obtain a patient-specific FE model from clinical MRIs through segmentation and reconstruction. Then, we will detail our latest FE breast model along with a new simulation and optimization pipeline only relying on a quick surface acquisition of the patient in the supine stance. Finally, we will investigate the importance of the breast's rheological properties and the significance of the infra-mammary ligament design on the overall simulation.
	
	\section{Method}
	\subsection{Data acquisition}
	Our study is an observational, retrospective study carried out by the gynecological surgery department of the Centre Hospitalier Universitaire (CHU) of Montpellier. The patient was selected from the``Programme de Médicalisation des Systèmes d’Information'' database. We obtained consent from the patient for the study and a favorable opinion from the``Comité Local d’éthique Recherche''\footnote{obtained on the $07/16/2017$ under the label $2017\_\texttt{CLER-MTP}\_07\texttt{-}04$. The study was declared in the registry of the CNIL (MR$003$) under the name of the ``Centre Hospitalier Universitaire'' (CHU) of Montpellier and registered on the ClinicalTrials website \url{https://www.clinicaltrials.gov/ct2/show/study/NCT03214419}.}. The patient has undergone MRI in prone configuration, breasts hanging downwards, arms abducted above the head, and a surface acquisition in the intra-operative (supine) position. It is important to note that the MRI is not specific to this study and is performed routinely as part of the preoperative workup for each patient. Therefore, the clinical pipeline can vary from one country or one hospital to another.
	
	\subsection{Image segmentation and reconstruction}
	Following the MRI (axial section, T2 injected sequence), a series of 2D grey-scale images are obtained, displaying the different organs of the patient. On each slice, the mammary gland, adipose tissues, pectoralis muscle, skin, and tumor were segmented. We used the open-source software 3D Slicer~\cite{Fedorov2012} to perform a semi-automatic segmentation, mixing thresholding, and growing regions methods~\cite{Song2017}. Then, 3D reconstruction was performed using the Marching Cubes~\cite{Lorensen1987} algorithm to obtain a 3D mesh describing the external surface of each organ. 
	
	\subsection{Finite element mesh}
	The reconstruction resulted in a mesh of more than \SI{308 000} triangles, which would lead to a 3D mesh of more than 1 billion volume elements. Billions of DOFs (Degree of Freedoms) are incompatible with clinical time scale without resorting to Machine Learning~\cite{Deshpande2022, Deshpande2022-2} or Model Order Reduction~\cite{Chinesta2014, Goury2016, Goury2018} techniques. One classical solution is to coarsen the original mesh to achieve fewer tetrahedra as an output of the Delaunay algorithm~\cite{Ito2015}. One major drawback of this method is the geometry accuracy loss compared with the original shape. Thus, we used the software InstantMesh~\cite{Jakob2015} to decimate and coarsen the original skin mesh while preserving the initial boundaries as much as possible. The result allowed to decrease the mesh size from \SI{308 844} triangles to only \SI{2040}. Finally, we used the software Gmsh~\cite{gmsh} to generate the volume mesh made of solely \SI{3025} tetrahedra. Note that alternative coarsening approaches were proposed in~\cite{Jacquemin2021} for strong form methods or a posteriori error estimators applied to biomechanical problems~\cite{Duflot2008, Bui2017, Duprez2020, Bulle2021}.
	
	\subsection{A novel breast finite element model}
	\label{sec:breast-model}
	The breast model we developed rests upon:
	
	\begin{itemize}
		\item A rigid pectoral muscle described by the domain $\Omega_{\mathrm{muscle}}$. Despite an evident deformable behavior of the pectoral muscle, this modeling hypothesis is also shared with several studies~\cite{Carter2009, Carter2012, Eiben2016}. ~\cite{Mira2018, BabarendaGamage2019} considered the deformation of the pectoral via FE models remarking its significant influence on the final breast shape. But for the sake of obtaining a simple and fast model, designing a rigid pectoral will drastically limit the complexity of the problem.
		
		\item A FE model of the breast volume embedding the adipose tissues, mammary gland, and tumor made of tetrahedra. The breast domain is denoted by $\Omega_{\mathrm{breast}}$ while the domain boundaries are indicated by $\Gamma_{\mathrm{breast}}$. The domain $\Gamma_{\mathrm{breast}}$ is subdivided into $3$ sub-domains $\Gamma_{\mathrm{breast}} = \Gamma_{\mathrm{inner}} \cup \Gamma_{\mathrm{skin}} \cup \Gamma_{\mathrm{ligament}}$.
		In this model, we assumed a single material embedding the glandular and adipose tissues. This hypothesis is relatively common in the literature~\cite{Carter2009, Carter2012, Mira2018, BabarendaGamage2019} as modeling heterogeneity could be tedious. Despite an evident non-linear, anisotropic, heterogeneous behavior, we modeled the breast using an isotropic Hooke's law coupled with corotational strains~\cite{Nesme2005}. A heterogeneous model could have handled more complex deformations and certainly better represent the actual breast behavior. Similarly to anisotropic behavior, this implies tuning additional parameters while increasing the overall uncertainty of the model. Finally, a non-linear material model could have expanded the deformation space but at the cost of solving a non-linear system, thereby increasing the computational cost. The equilibrium equations for the motion of the breast read: 
		\begin{equation}
			\begin{aligned}
				-\nabla \cdot \boldsymbol{\sigma} &=\boldsymbol{f} \text { in } \Omega_{\mathrm{breast}}, \\
				\boldsymbol{\sigma} &=\lambda \operatorname{tr}(\boldsymbol{\varepsilon}) \boldsymbol{I}+2 \mu \boldsymbol{\varepsilon}, \\
				\boldsymbol{J} &=\boldsymbol{R}_{q r} \cdot  \boldsymbol{\varepsilon}, \\
				\boldsymbol{\varepsilon} &=\boldsymbol{R}_{q r}^{-1} \boldsymbol{J}=\left[\begin{array}{ccc}
					1+\boldsymbol{\varepsilon}_{x x} & 2 \boldsymbol{\varepsilon}_{x y} & 2 \boldsymbol{\varepsilon}_{x z} \\
					0 & 1+\boldsymbol{\varepsilon}_{y y} & 2 \boldsymbol{\varepsilon}_{y z}. \\
					0 & 0 & 1+\boldsymbol{\varepsilon}_{z z}
				\end{array}\right]
			\end{aligned}
		\end{equation}
		Where $\boldsymbol{\sigma}$ is the Cauchy stress tensor, $\boldsymbol{f}$ are the body forces per unit of volume, $\mu$ and $\lambda$ are Lamé elastic parameters representing the material in $\Omega_{\mathrm{breast}}$, $\boldsymbol{I}$ is the identity tensor, $\operatorname{tr(\bullet)}$ is the trace operator, $\boldsymbol{J}$ is the displacement matrix decomposed with the $QR$ method in order to extract separately a rigid rotation $\boldsymbol{R}_{q r}$ and the deformation matrix $\varepsilon$.
		
		\item A FE model of the skin on $\Gamma_{\mathrm{skin}}$ made of 3D triangles. In the same manner, as the breast, we used the same set of equations for the $\Omega_{\mathrm{skin}}$ domain using different Lamé parameters. The Lamé coefficients can be expressed as a function of the Young's modulus ($E$): $\lambda = \frac{E \nu}{(1+\nu)(1-2 \nu)}$ and Poisson's ratio ($\nu$): $\mu = \frac{E}{2(1+\nu)}$. An incompressible isotropic material is usually modeled with a Poisson's ratio of exactly $0.5$. This hypothesis has been commonly used for breast simulations, especially for breast compression simulations. In this study, to avoid additional numerical complexity during the optimization phase, we considered a nearly incompressible breast by imposing $\nu_{\mathrm{breast}} =\nu_{\mathrm{skin}} = 0.45$, while $E_{\mathrm{breast}}$ and $E_{\mathrm{skin}}$ will be the optimized parameters.
		
		\item A 1D ligament attaching the breast to the pectoral muscle. In addition to the breast,~\cite{McGhee2020} describes a deep ligament between the inner part of the breast boundaries and the pectoral muscle (figure \ref{fig:ligament}). For each DOF of the ligament (belonging to $\Gamma_{\mathrm{ligament}}$), we used the ICP (Iterative Closest Points) algorithm~\cite{Besl1992} to find the closest point on the domain $\Omega_{\mathrm{muscle}}$. Hence, we prescribed for each DOFs a constraint violation where the intensity of the forces is defined by $\lambda = -\frac{1}{c} ( \boldsymbol{x} - d \boldsymbol{v})$. Where $\boldsymbol{x}$ and $\boldsymbol{v}$ are the positions and velocities of the DOFs. $c$ and $d$ are the respective compliance and damping ratio. The damping ratio is fixed to $0$ and the compliance to $10^{-8}$ \si{m/N}, enforcing a strong attachment between the breast and the pectoral on the ligament boundaries.
		
		\item Sliding boundary conditions between the breast and the pectoral muscle. The mechanical interface between the breast and the pectoral muscle has been poorly studied. Obtaining physically relevant quantities from the literature, such as contact and friction laws (if relevant), is a problematic task. Even if available, the contact formulation computation would negatively impact the run-time and stability of our rapid simulation. To replicate the boundary condition between the pectoral and the breast, we again used the ICP to find the closest point from the inner breast surface ($\Gamma_{\mathrm{inner}}$) to the pectoral muscle ($\Gamma_{\mathrm{muscle}}$). Once the correspondence was established, we used a stable compliant formulation described in~\cite{Tournier2017} to mimic the sliding between the $2$ surfaces.
	\end{itemize}
	
	\subsection{Simulation pipeline}
\begin{figure}[h]
		\centering
		\includegraphics[width=\textwidth]{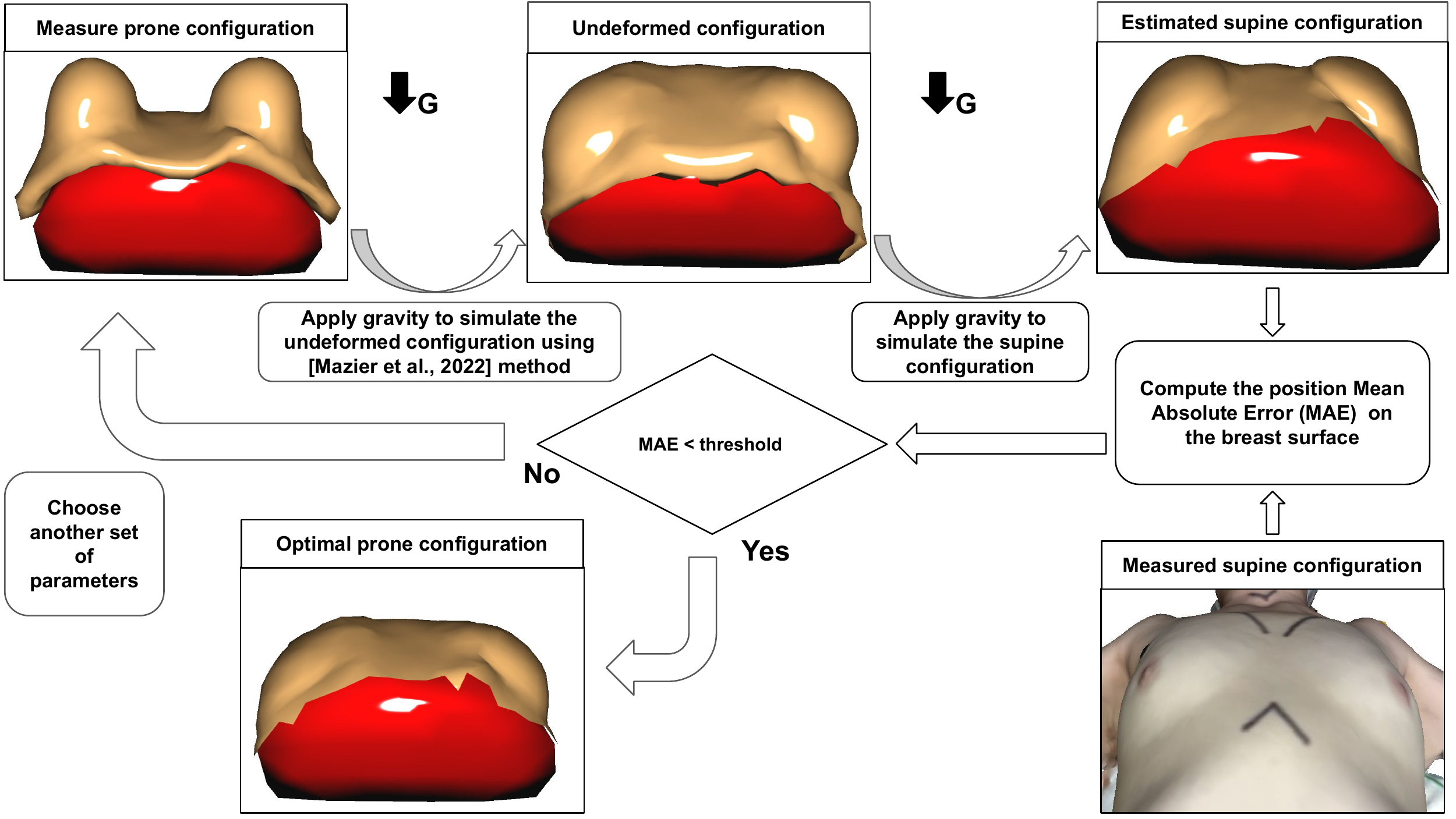}
		\caption{Simulation pipeline for estimating the patient-specific intra-operative (supine) configuration. The process starts in the top left corner, where the measured prone configuration geometry is obtained from the MRI by segmentation and 3D reconstruction. Then, the theoretical undeformed configuration is obtained using the inverse method described in~\cite{Mazier2022} by applying gravity in the opposite direction of the sagging breast. Afterward, gravity is again applied to estimate the final intra-operative pose. The estimation is compared with the measured intra-operative configuration obtained with a surface acquisition device (assumed to be the gold standard) by assessing the Mean Absolute Error (MAE defined in equation \ref{eq:MAE_metric}) on the breast surface. If the MAE is superior to the user-defined threshold, the pipeline is run again, choosing a different set of mechanical parameters. Otherwise, the estimated supine configuration is sufficiently close to the patient's intra-operative pose. Image inspired from~\cite{Mira2018}.}
		\label{fig:simulation-pipeline}
	\end{figure}

	We chose the Simulation Open Framework Architecture (SOFA ~\cite{Faure2012}) to perform our simulations. SOFA allows for running real-time rendered simulations, matching the central goal of our work. We followed a similar simulation pipeline as~\cite{Mira2018}, described in figure \ref{fig:simulation-pipeline}. Namely, we obtained the theoretical undeformed configuration using an inverse method described in~\cite{Mazier2022} by applying gravity in the opposite direction to the sagging breast in the prone configuration. Afterward, gravity is again applied to estimate the final intra-operative pose. We finally compared the estimated solution with the measured intra-operative configuration obtained with a surface acquisition device (assumed to be our gold standard). The comparison was assessed by calculating the Mean Absolute Error (MAE) on the breast surface given by
	
	\begin{equation}
		\label{eq:MAE_metric}
		\textrm{MAE}= \frac{1}{N_{CP}} \sum_{i=1}^{N_{CP}}\left|d_{i}\right| =\sum_{i}^{N_{CP}} | \mathcal{P}_i (M(\boldsymbol{x}, E_{\mathrm{breast}}, E_{\mathrm{skin}}, \nu_{\mathrm{breast}}, \nu_{\mathrm{skin}}), V) |.
	\end{equation}
	$N_{CP}$ stands for the number of closest points between the model prediction $M$ and the measured configuration $V$ (surface acquisition), while $| \bullet |$ denotes the absolute value operator. $\mathcal{P}_i$ is an operator projecting the model points ($M$) onto the closest primitive (triangle, edge, or point) of $V$ resulting in the distance $d_i$. This allows for a more accurate registration rather than point-to-point distances, and differentiating \ref{eq:MAE_metric} with respect to the points of $M$ can be easily done by computing the normal vectors of the triangles of $V$. 
	
	If the MAE exceeds a selected threshold, the pipeline is run again, choosing a different set of mechanical parameters. Otherwise, the estimated supine configuration is considered sufficiently close to the patient's intra-operative pose.
	
	Several differences can be noted compared with~\cite{Mira2018}. First, \cite{Mira2018} simulated the imaging configuration from the intra-operative stance, while we simulated the opposite path. Secondly, \cite{Mira2018} optimized the mechanical properties based on the simulation from undeformed to imaging configuration without recomputing the undeformed stance. We chose a different approach. After evaluating the difference between the simulated and measured intra-operative pose, we restart the complete pipeline from the imaging to the intra-operative estimation. It allows us to optimize the material parameters [$E_{\mathrm{breast}}$, $E_{\mathrm{skin}}$] on the complete pipeline and to adjust the undeformed configuration accordingly. One reason explaining this difference could be the algorithm used to infer the undeformed pose (~\cite{Govindjee1998} for ~\cite{Mira2018} compared with~\cite{Mazier2022} in this pipeline).
	
	\subsection{Optimization}
	\label{sec:optimization}
	Classical gradient-descent methods are commonly used for optimization procedures. However, they possess two main drawbacks. First, they require access to the first (sometimes also the second) derivative of the cost function with respect to the (mechanical) parameters. Secondly, they strongly depend on initial values and frequently fail when encountering non-convex or rugged search landscapes (e.g., sharp bends, discontinuities, outliers, noise, and local optima). To alleviate those hurdles, statistical optimizers, such as the Covariance Matrix Adaptation Evolution Strategy (CMA-ES), have proven to be also reliable. CMA-ES employs an evolutionary strategy using a set of $\mu$ parents to produce $\lambda$ children~\cite{Auger2005}. Each child is produced by recombining $\rho$ parents and is usually mutated by adding a random variable following a normal distribution. Through the iterations, the algorithm relies on the variance-covariance matrix adaptation of the multi-normal distribution used for the mutation. The complete methodology of the algorithm are explained in~\cite{Hansen2016} and recent applications of a similar algorithm are shown in~\cite{Jansari2022}.
	
	\section{Results}
	\subsection{Influence of the optimizer}
	\label{sec:optimization_influence}
\begin{figure}[h]
		\centering
		\begin{subfigure}[t]{0.5\textwidth}
			\centering
			\includegraphics[width=\textwidth]{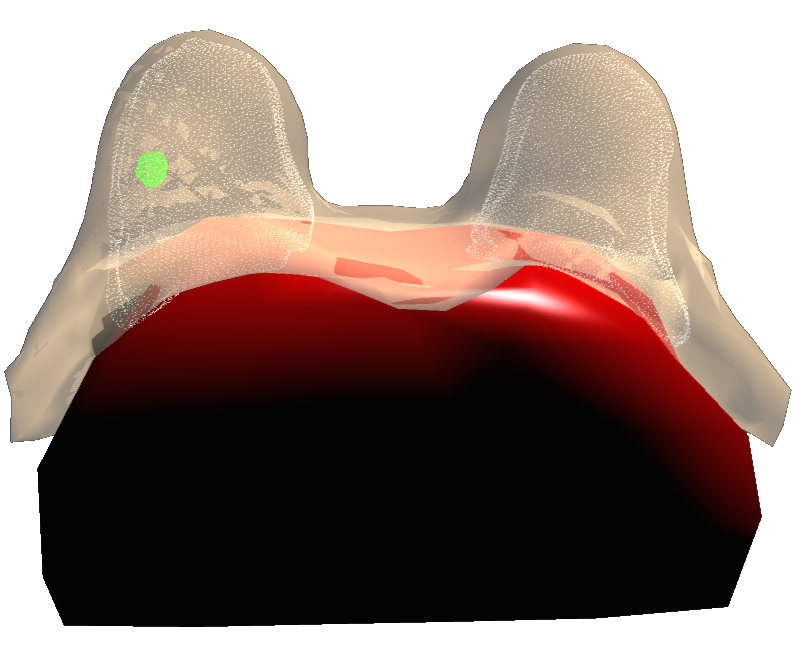}
			\caption{}
		\end{subfigure}%
		~ 
		\begin{subfigure}[t]{0.5\textwidth}
			\centering
			\includegraphics[width=\textwidth]{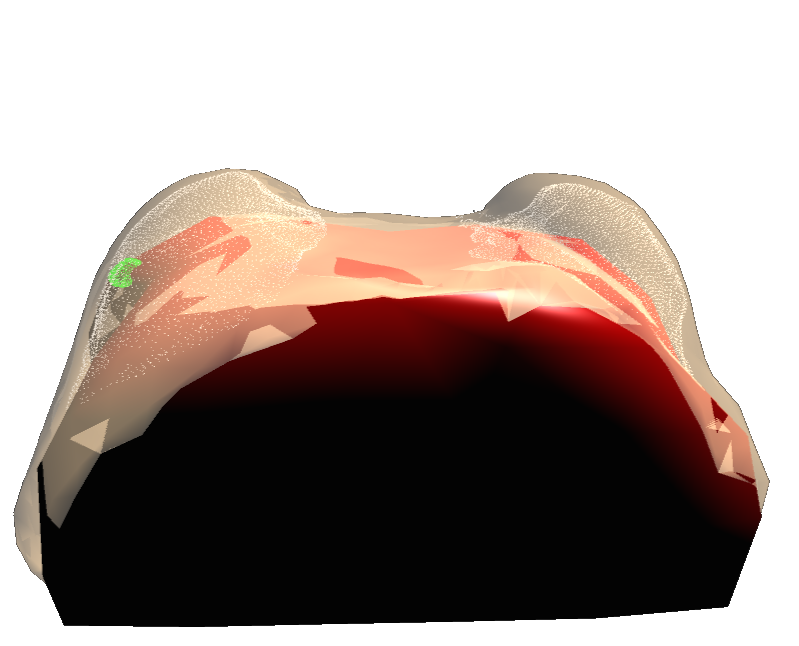}
			\caption{}
		\end{subfigure}
		\caption{Complete simulation including the skin, adipose tissues (in yellow transparent), the mammary gland (white dots), and the tumor (green dots). (a) Imaging configuration of the patient (prone). (b) Estimated patient-specific intra-operative configuration (supine) after optimization of the mechanical properties.}
		\label{fig:finale}
	\end{figure}

	The simulation (including the $2$ simulations, namely, from the prone to undeformed stance and from the undeformed to the supine configuration) takes less than \SI{20}{s} and is depicted in figure \ref{fig:finale}. In table \ref{table}, we summarized various initial conditions and standard deviations, resulting in slightly different optimized mechanical properties (and the number of iterations needed). Due to the statistical nature of the algorithm, an average of the output has been performed over $10$ sequences.
	
	\begin{table}[h]
		\centering
		\begin{tabular}{||c c c c c c c c||} 
			\hline
			$E^{\mathrm{init}}_{\mathrm{breast}}$ & $E^{\mathrm{init}}_{\mathrm{skin}}$ & $\sigma^{\mathrm{init}}_{\mathrm{breast}}$ & $\sigma^{\mathrm{init}}_{\mathrm{skin}}$ & $E^{\mathrm{opti}}_{\mathrm{breast}}$& $E^{\mathrm{opti}}_{\mathrm{skin}}$ & MAE [mm] &\# iter \\ [0.5ex] 
			\hline\hline
			0.3 & 75 & 0.1 & 50 & 0.26 & 20.50 & 4.17 & 15\\ 
			\hline
			0.3 & 75 & 0.3 & 100 & 0.32 & 22.72 & 4.00 & 30\\
			\hline
			1.5 & 120 & 1.5 & 100 & 0.35 & 30.83 & 4.02 & 101\\
			\hline
		\end{tabular}
		\caption{CMA-ES (Covariance Matrix Adaptation Evolution Strategy) optimization of $E_{\mathrm{breast}}$ and $E_{\mathrm{skin}}$ to minimize the Mean Absolute Error (MAE) defined in equation \ref{eq:MAE_metric}. Different initial values of the parameters ($E^{\mathrm{init}}_{\mathrm{breast}}$, $E^{\mathrm{init}}_{\mathrm{skin}}$) in [\si{kPa}] and standard deviations ($\sigma^{\mathrm{init}}_{\mathrm{breast}}$, $\sigma^{\mathrm{init}}_{\mathrm{skin}}$) in [\si{kPa}] have been selected. The results averaged over $10$ sequences are the optimized Young's modulus of the breast and skin $E^{\mathrm{optimized}}_{\mathrm{breast}}$, $E^{\mathrm{optimized}}_{\mathrm{skin}}$ in [\si{kPa}] and the number of iterations (\# iter) needed to minimize the MAE in [\si{mm}]. In the first row, mechanical parameters were initialized according to the literature with a reduced standard deviation (indicating high confidence in the initial parameters). In the second row, while fixing the initial parameters, the initial standard deviations were expanded (indicating less confidence in the initial parameters). On the last line, both initial parameters and standard deviation were increased (indicating no prior knowledge of the mechanical properties).}
		\label{table}
	\end{table}
	
	\subsection{Influence of the rheological parameters}
	\label{sec:sensitivity}
	In the previous section, we have shown that several values of $E_{\mathrm{breast}}$ and $E_{\mathrm{skin}}$ could equally minimize the error between the estimated and measured supine configuration. Hence, quantifying the influence of mechanical properties on the final breast shape appears necessary. Consequently, we used Monte Carlo (MC) simulations which rely on a stochastic distribution to understand the parameters' influence on the simulation~\cite{Hauseux2017, Hauseux2018}. Note that Bayesian inference could also have been used for uncertainty quantification in biomechanics~\cite{Rappel2019, Rappel2019-2, Rappel2020}. Without any prior information on the two parameters $E_{\mathrm{breast}}$ and $E_{\mathrm{skin}}$, we assumed that they were both following a normal distribution with a mean value $\overline{E}$ and standard deviation $\sigma$
	
	\begin{align}
		E_{\mathrm{breast}} \sim N(\overline{E_{\mathrm{breast}}}, \sigma_{\mathrm{breast}}^{2}), \\ 
		E_{\mathrm{skin}} \sim N(\overline{E_{\mathrm{skin}}}, \sigma_{\mathrm{skin}}^{2}).
	\end{align}
	
\begin{figure}[h]
		\centering
		\includegraphics[width=\textwidth]{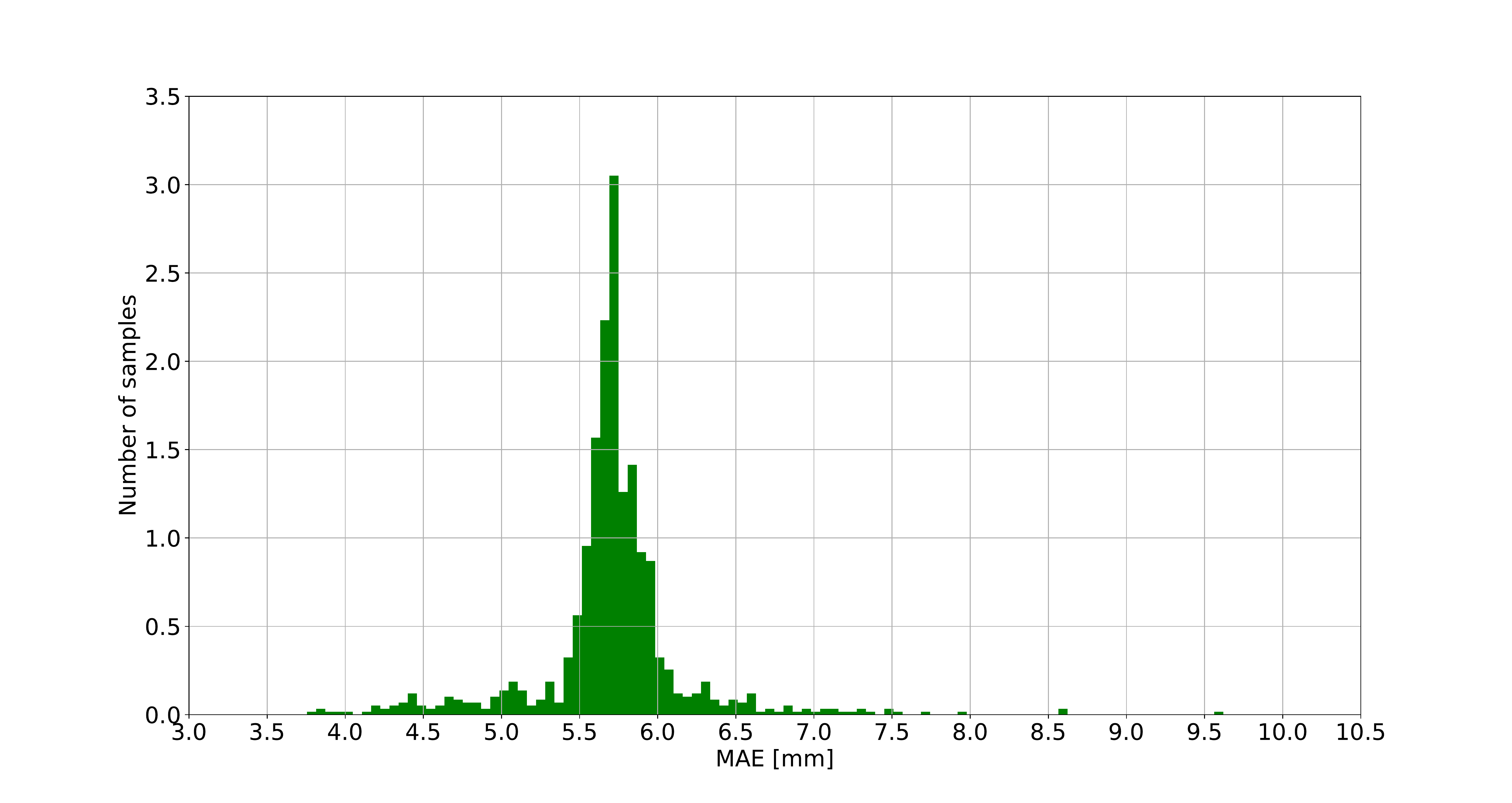}
		\caption{Mean Absolute Error (MAE) statistics of the $1000$ Monte Carlo (MC) simulations using $2$ normal distributions of the mechanical parameters $E_{\mathrm{breast}} \sim N(0.32, 0.4)$ [\si{kPa}] and $ E_{\mathrm{skin}} \sim N(23, 40)$ [\si{kPa}]~\cite{Hauseux2017, Hauseux2018, Rappel2019, Rappel2019-2, Rappel2020}.}
		\label{fig:mc_stats}
	\end{figure}

	We chose $\overline{E_{\mathrm{breast}}} = 0.32$ and $\overline{E_{\mathrm{skin}}} = 23$ \si{kPa} as they were close to the previous optimal solution. To slightly perturb the parameters, we chose the following standard deviations $ \sigma_{\mathrm{breast}}^{2} = 0.4$ and $\sigma_{\mathrm{skin}}^{2} = 40$. The standard deviations are large enough to allow considerable variations of the parameter space. We drew $1000$ samples of each parameter and ran the simulation with the stochastic parameter sets, and the results are presented in figure \ref{fig:mc_stats}.
	
	\subsection{Influence of the circum-mammary ligament}
	\label{sec:ligament}
\begin{figure}
		\centering
		\begin{subfigure}[t]{0.5\textwidth}
			\centering
			\includegraphics[width=\textwidth]{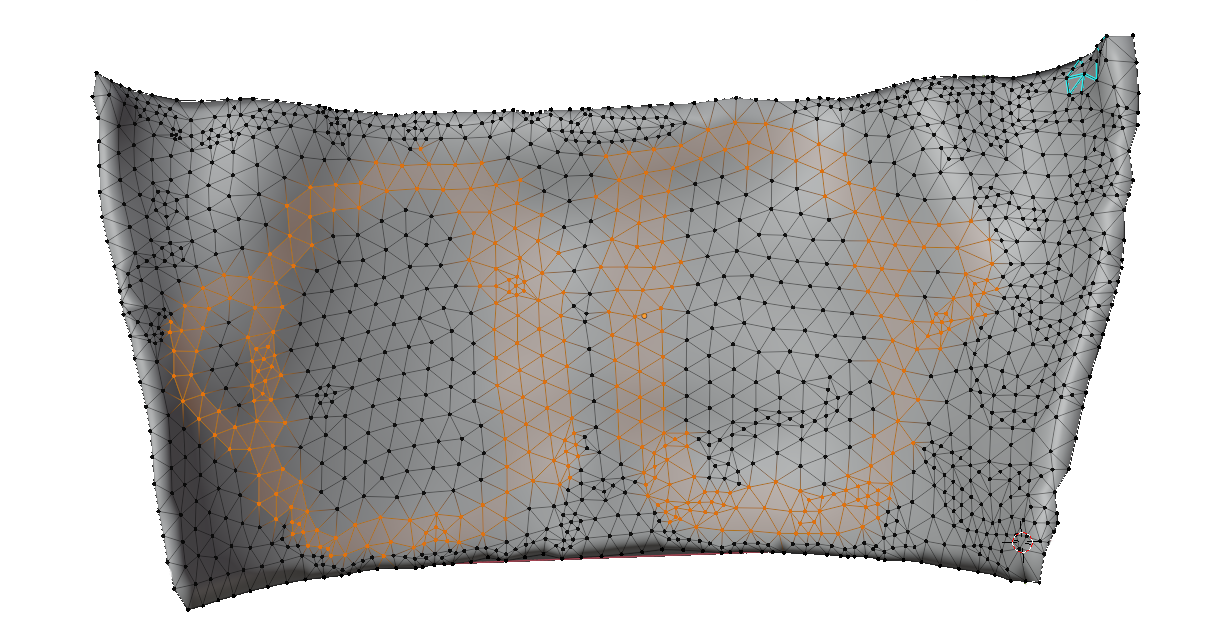}
			\caption{}
		\end{subfigure}%
		~ 
		\begin{subfigure}[t]{0.5\textwidth}
			\centering
			\includegraphics[width=\textwidth]{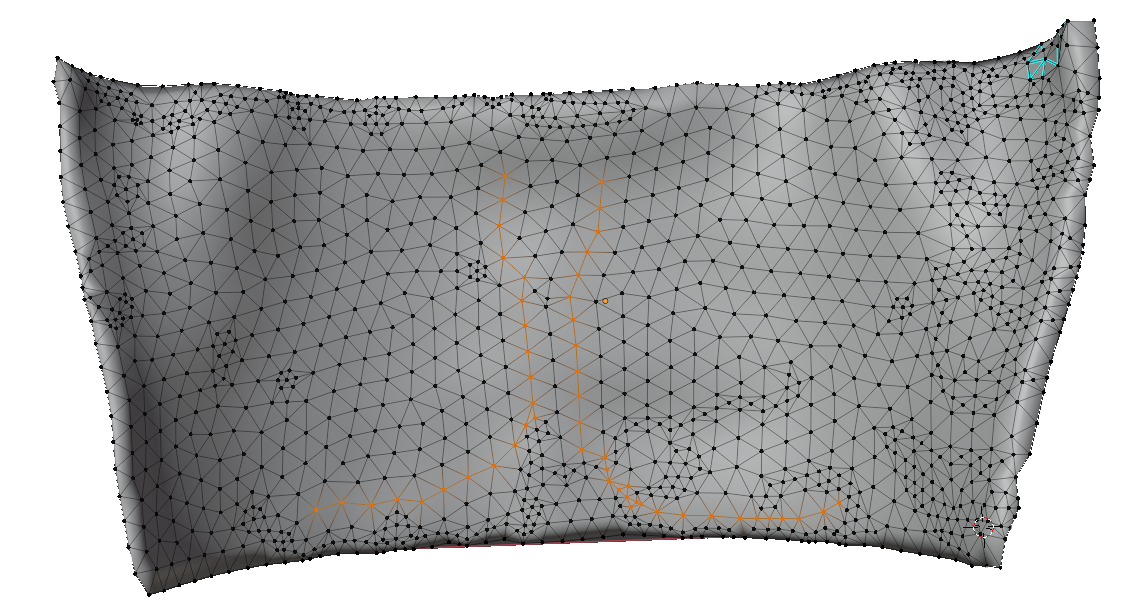}
			\caption{}
		\end{subfigure}
		\begin{subfigure}[t]{\textwidth}
			\centering
			\includegraphics[width=\textwidth]{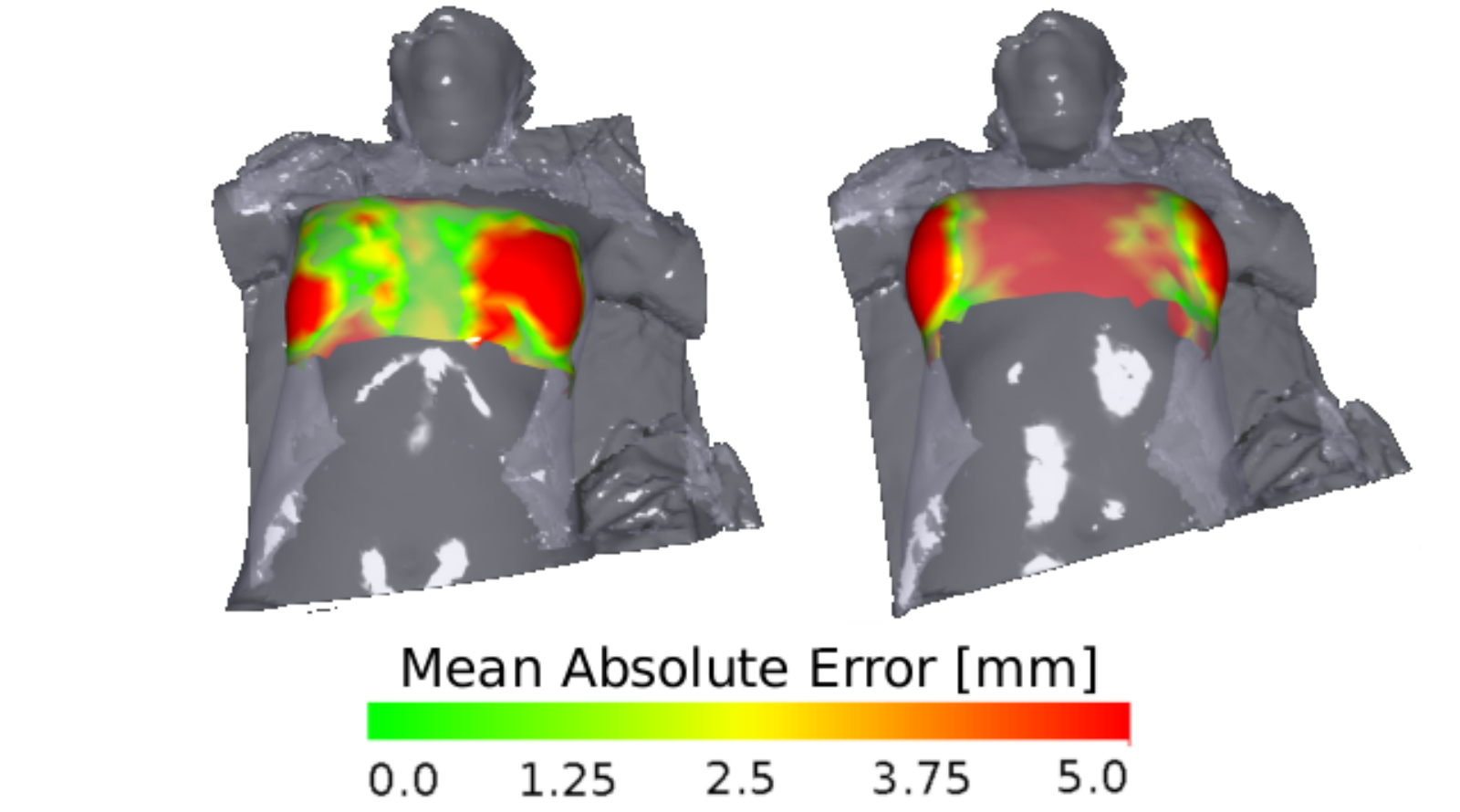}
		\end{subfigure}
		\caption{Impact of the ligament stiffness and position on the simulation. The nodes of the inner breast side in orange delineate the infra-mammary ligament geometry. (a) Complete ligament description as described in \cite{McGhee2020}. (b) Altered thinner ligament description. On the bottom, the simulation result from the prone to the supine configuration with optimized mechanical properties. The color bar displays the Mean Absolute Error (MAE) between the estimated and measured (in grey) prone configuration. On the bottom left, the simulation result with the complete ligament design (corresponding to (a)) in an MAE of $4.00$ \si{mm}. On the bottom right, the simulation result with the thinner altered ligament design (corresponding to (b)) in an MAE of $8.24$ \si{mm}.}
		\label{fig:ligament}
	\end{figure}

	We previously showed that the breast shape was slightly sensitive to the rheological parameters of the model. In this section, we additionally study the impact of the circum-mammary ligament design on the simulation. This ligament is described in~\cite{McGhee2020} and depicted as a deep connection between the breast and the pectoral muscle. To observe the influence of the ligament, we fixed the optimal mechanical properties of the breast and only altered the thickness and position of the original ligaments (figure \ref{fig:ligament} (a)) to a thinner one (figure \ref{fig:ligament} (b)). While obtaining an MAE of $4.00$ \si{mm}, the altered ligament resulted in an MAE of $8.24$ \si{mm}, doubling the error compared with the proper ligament geometry.
	
	\section{Discussion}
	In this study, we show that our pipeline can estimate patient-specific breast deformations from imaging to intra-operative configuration within clinical time and without perturbing the surgical procedure. In section \ref{sec:optimization_influence}, we first investigated the influence of CMA-ES optimizer initial parameters. From table \ref{table}, several points are notable.
	
	Firstly, the optimized values of $E^{\mathrm{opti}}_{\mathrm{breast}}$, $E^{\mathrm{opti}}_{\mathrm{skin}}$ are consistent with the literature. For the breast elasticity, similarly to~\cite{Mira2018}, we obtain a value between $0.26$ and $0.35$ \si{kPa}, belonging to the low range of the breast Young's modulus [$0.3$ \si{kPa}: $6$ \si{kPa}] given in the literature. For skin elasticity, we obtain a wide range of optimized Young's moduli between $20.50$ and $30.83$ \si{kPa}. According to the literature, the range of skin Young's modulus is [$7.4$ \si{kPa}: $58.4$ \si{kPa}], placing our optimized values in the middle range.
	
	Secondly, by trying different initial values and standard deviations, we tested the sensitivity of the CMA-ES optimizer to the initial conditions. For the given MAE, CMA-ES appears to converge to different optimal configurations depending on the initial and standard deviation of the parameters. Indeed, between the first two rows of table \ref{table}, we kept the same initial parameters (close to the literature) but increased the initial standard deviation (indicating lower confidence in the initial parameters). The results of choosing more significant initial standard deviations are an increase in the number of iterations (factor of 2 higher in this case), stiffer optimized parameters, and a decrease of \SI{0.1}{mm} in the MAE. Obtaining more iterations is expected as expanding the initial standard deviation values increases the search domain, thus demanding more time to converge. However, augmenting the search domain enabled obtaining a lower MAE value. This indicates that too narrow initial standard deviation values could considerably restrict the search space and ignore potential optimal solutions. Although, gradient-descent optimizers could encounter complications using initialization which are too distant from the optimized solutions. On the third row of table \ref{table}, we drastically increased the initial values of the mechanical properties. This led to a severe increase in the number of iterations (101 compared to 15 or 30) and similar values of the MAE but, again, slightly different values of the optimized parameters. This indicates that despite a flawed initialization, the CMA-ES optimizer can still converge to acceptable solutions, but multiple combinations of parameters can lead to comparable error measures. 
	
	Thirdly, we did not reach an MAE lower than \SI{4}{mm}. Compared with the literature, the error is acceptable $1.90$ $\pm$ \SI{2.17}{mm} to \SI{5}{mm} using mean and RMS errors. However, we expect more complex material laws such as Mooney-Rivlin or Neo-Hooke materials could decrease the MAE by better fitting the surface acquisition. This would indicate that the corotational strain is may not be sufficient to model such large deformations, and a more classical deformation measure may be more suitable.
	
	In section \ref{sec:sensitivity}, figure \ref{fig:mc_stats} suggests that the mechanical parameters have a moderate impact on the resulting shape of the breast. Indeed, by taking extremes values of the normal distributions e.g. $E_{\mathrm{breast}} = 0.8 $ \si{kPa} and $E_{\mathrm{skin}} = 80$ \si{kPa}, the MAE is equal to $6.56$ \si{mm} approximately $2.5$ \si{mm} off the optimal MAE. We obtain a Gaussian distribution of mean \SI{5.70}{mm} and a standard deviation of \SI{0.48}{mm}. Hence, the standard deviation of the MAE distribution indicates that a respective error of $\pm$ 40 \si{kPA} or $\pm$ 0.4 \si{kPA} for the stiffness of the skin and breast could have a 68 \% of chance of creating an error less than \SI{0.48}{mm}. Consequently, the parameter identification enables a significant decrease in the MAE, but a slight error in the parameters could still lead to an acceptable MAE (as long as physical parameters are obtained).
	
	In section \ref{sec:ligament}, figure \ref{fig:ligament} indicates that designing a thinner ligament (figure \ref{fig:ligament}a) without describing a full circle (figure (\ref{fig:ligament}b) to sustain the inner breast leads to a higher MAE. By conserving the same optimal material properties and altering only the ligament design, the MAE rose from $4.00$ to $8.24$ \si{mm}, especially close to the sternal region and breasts. Indeed, the ligament is not constraining the breast as before (almost fixed to the sternum), resulting in an underconstrained breast which deforms too much on the sides.
	
	Despite converging to an MAE of \SI{4}{mm}, it is impossible to validate the tumor position with the current dataset. Indeed, surface acquisitions only allow measuring the visible deformations, and volume acquisitions such as MRIs are required for internal validation. Therefore, MRIs in the intra-operative configuration are clinically irrelevant for surgeons, and another investigation is needed to validate our model. 
	
	This study is still in its early stages, and more validations and improvement steps could be implemented. A deeper study could focus on the efficiency of gradient-descent algorithms or other gradient-free methods. Indeed, for the moment, the CMA-ES optimizer is slow to converge (especially if initial parameters are distant from the optimal ones), making practical use challenging. Further, as shown in figure \ref{fig:ligament}, the MAE is lower on the left breast compared with the right breast. This indicates that choosing different material properties for the left and right breast (as in~\cite{Mira2018}) could be relevant. In addition, we designed the breast using a linear elastic model to describe the stress-strain relation. The co-rotational model allowed sufficient deformations to reach the supine configuration. However, more complex material models such as Mooney-Rivlin, Yeoh, and Gent~\cite{Mira2018-2, Elouneg2021, Sutula2020} enabling a natural stiffening effect, characteristic of collagen fibers, could be effortlessly tested in SOFA~\cite{sonics}. Finally, the muscle was assumed to be a rigid body, while~\cite{BabarendaGamage2019} demonstrated that neglecting the pectoral deformation during the repositioning of the volunteer may strongly impact the estimates in prone and supine tilted breast configurations. At the same time, such complications of the model would also increase the number of parameters. The open questions remain: ``What is the simplest material law which could be used to model breasts as a whole?~\cite{Elmukashfi2022} Would composite breast models enable a more faithful representation of experimental and clinical results, and at which costs? Could intra-operative data simplify this model by transferring information to which even simplistic models could be fitted~\cite{Plantefve2015, Nikolaev2020}?''
	
	\section{Conclusion}
	To conclude, we presented a complete pipeline from the medical image to an optimized, simple, and patient-specific breast model. We first demonstrate the manual process to obtain the patient's biomechanical model from the MRI in the prone pose. Then, using the CMA-ES optimizers, we were able to retrieve the personalized mechanical properties of the patient's breast by solely using a surface acquisition. Hence, we studied the sensitivity of the optimizer to the initial parameters, the rheological properties, and the design of the circum-mammary ligament in the simulation process. We were finally able to include the mammary gland and tumor in the simulation but couldn't evaluate the accuracy of our prediction due to a lack of data. 
	The final results are an MAE of \SI{4.00}{mm} for the mechanical parameters $E_{\mathrm{breast}} = 0.32$ \si{kPa} and $E_{\mathrm{skin}} = 22.72$ \si{kPa}. The mechanical parameters are congruent with the literature found in~\cite{Mira2018, BabarendaGamage2019}. The simulation (including finding the undeformed and prone configuration) takes less than \SI{20}{s}. The CMA-ES optimizer converges, on average, between $5$ to \SI{30}{min} depending on the initialization of the parameters. The final Mean Absolute Error is slightly greater to the one found by~\cite{Mira2018}, but to our knowledge, our model is the fastest to converge to such a low MAE.
	
	\section*{Acknowledgements}
	The medical images used in the present study were obtained from Hopital Arnaud de Villeneuve, Département de Gynécologie Obstétrique in collaboration with Dr. Gauthier Rathat. The authors would like to acknowledge AnatoScope for their help in the modeling process and for accessing their private SOFA version. The authors acknowledge Prof. Yohan Payan and Dr. Thiranja Prasad Babarenda Gamage for interesting discussions. Stéphane Bordas thanks the visiting position at China Medical University, Taiwan. This project has received funding from the European Union’s Horizon 2020 research and innovation programme under the Marie Sklodowska-Curie grant agreement No. 764644. This publication only contains the RAINBOW consortium’s views and the Research Executive Agency and the Commission are not responsible for any use that may be made of the information it contains. The authors thank the support of the FNR and ANR grant S-Keloid No. 16399490.
	
	\bibliography{mybibfile}
	
\end{document}